\documentclass[aps,twocolumn,groupedaddress,showpacs]{revtex4-1} 
\usepackage{graphicx,amsmath}
\usepackage{color}
\usepackage{bm}

\begin{document}

 
\title{Coarse-grained description of general oscillator networks} 


\author{Yuki Izumida}
\email[]{izumida.yuki@ocha.ac.jp}
\author{Hiroshi Kori}
\email[]{kori.hiroshi@ocha.ac.jp}
\affiliation{Department of Information Sciences, 
Ochanomizu University 2-1-1 Ohtsuka, Bunkyo-ku, Tokyo 112-8610, Japan}



\begin{abstract}
We propose a novel and systematic method for coarse-graining oscillator networks described by phase equations.
Our coarse-graining method enables us to obtain the closed coarse-grained equations for a few effective eigenmodes, 
which is based on the eigenvalue problem of the linearized system around the phase-locked solution and a nonlinear transformation.
We demonstrate our method by applying it to oscillator dynamics on a random graph, which exhibits a saddle-node bifurcation at the bifurcation point.
\end{abstract} 


\maketitle

\section{Introduction}
In natural or social phenomena emerging from complicated interactions between the elements, it is common that only a few variables, compared to other many degrees of freedom, play a vital role in the phenomena. In physics, elucidating these important variables and deriving their coarse-grained dynamical equations are especially useful for understanding the essence of these phenomena.

A coarse-grained description is particularly important in dynamical processes on complex networks~\cite{RK}.
To highlight the essential nodes in networks, many types of centrality measures have been proposed~\cite{BA}. However,
it may be possible that a group of nodes instead of individual nodes collectively dominate the dynamics on the network. 
A coarse-grained description
would help us to identify such a combination of nodes.

A collection of oscillators interacting with each other (oscillator networks)
shows a variety of collective behaviors: collective flushing by fireflies~\cite{PRK}, organized bursting of neural cells~\cite{PRK}, and quorum sensing by social amoebae~\cite{GFMS}, to name a few.
These oscillator networks may be described by phase equations~\cite{W,K} 
to which such a coarse-grained view could be applied. 
For phase-locked (synchronized) oscillators, 
only one eigenmode corresponding to the rotational invariance in the phase equations 
may become the unique effective degree of freedom~\cite{KKNAK}. 
Other eigenmodes correspond to the fluctuations of each oscillator around the phase-locked solution and thus are not important. 
On the other hand, other eigenmodes also become effective
when the system exhibits a bifurcation of the phase-locked solution at a bifurcation point, 
and they may exhibit more complicated collective behaviors other than synchronization.
Therefore, it is important to derive the coarse-grained dynamical equations for these effective eigenmodes closed only by them, 
which can be applicable even after the bifurcation of the phase-locked solution.

In this paper, we propose a novel and systematic method for coarse-graining oscillator networks described by phase equations. 
Our method enables us to obtain the closed coarse-grained equations for a few effective eigenmodes, 
which is based on the eigenvalue problem of the linearized system around the phase-locked solution and a nonlinear transformation.
Our method can be applied to the phase equations of any network structure with any system size, assuming they have a stable phase-locked solution.
We demonstrate our theory by applying it to an oscillator network on a random graph, which exhibits a saddle-node (SN) bifurcation of the phase-locked solution at the bifurcation point.

\section{Model}
We consider the following phase equations describing the time evolution of $N$ oscillators ${\bm \phi}=(\phi_1, \phi_2, \cdots, \phi_N)^{\rm T}$ in a network (oscillator network) 
in the presence or the absence of external forcing:
\begin{eqnarray}
\dot{\phi_i}=\omega_i+\sum_{j=1}^{N}\Gamma_{ij}^K(\phi_i-\phi_j)+\epsilon Z(\phi_i)\xi_i(t),\label{eq.phase_eq}
\end{eqnarray}
where $\phi_i$, $\omega_i$, and $\Gamma_{ij}^K$ denote the phase of the $i$th oscillator, 
its natural frequency, and the $2\pi$-periodic coupling function from the $j$th 
oscillator to the $i$th oscillator, respectively. Although we may regard $K$ as any parameter included in the coupling function,
we assume that $K$ is the overall coupling strength hereafter.
The functions $\xi_i(t)$ and $Z(\phi_i)$ represent the time-dependent external forcing 
acting on the $i$th oscillator and the phase sensitivity of the $i$th oscillator to the forcing, respectively.
The parameter $\epsilon$ represents the intensity of the external forcing. We set $\epsilon=0$ when the external forcing is absent.

\section{Coarse-graining method}
We propose a coarse-graining method for the oscillator networks described by the phase equations in Eq.~(\ref{eq.phase_eq}).
Our method consists of three steps: (A) nonlinear transformation of variables from the phases to the nonlinear eigenmodes,
(B) derivation of the dynamical equations for the nonlinear eigenmodes, and 
(C) coarse-graining by neglecting unimportant nonlinear eigenmodes, as we explain below.
\subsection{Nonlinear transformation of variables}
Here, we consider a variable transformation that enables us to reduce
Eq.~(\ref{eq.phase_eq}) to a system with a few essential variables.
The transformation is nonlinear, although the transformed variables asymptotically agree with the linear eigenmodes that diagonalize the system in the vicinity of the phase-locked solution.
This nonlinear transformation preserves the important invariance that Eq.~(\ref{eq.phase_eq}) has; i.e., 
the system is invariant under $\phi_i \to \phi_i + 2\pi$ for any $i$, while the linear transformation violates this invariance.
The reason why we want to preserve this invariance is to describe the dynamical behavior after the bifurcation of the phase-locked solution, where the system does not remain in the vicinity of the phase-locked solution anymore.

As a necessary condition for our theory to be applied, 
we assume the existence of a stable phase-locked solution of Eq.~(\ref{eq.phase_eq}) 
in the absence of the external forcing $\epsilon=0$ within a certain range of the coupling strength $K$ included in the coupling function. 
When $K$ is sufficiently large, the system is assumed to be in a phase-locked state.
As $K$ decreases, the phase-locked solution bifurcates at a bifurcation point $K_{\rm c}$, and more complicated collective behaviors can occur. 
Whereas we consider systems that exhibit a bifurcation of co-dimension one at the bifurcation point for simplicity, 
an extension of our method to systems admitting a bifurcation of a higher co-dimension is straightforward. 

We express such a stable phase-locked solution as 
\begin{eqnarray}
{\bm \phi}^*(t)=(\Omega t+\phi_1^0, \Omega t+\phi_2^0, \cdots, \Omega t+\phi_N^0)^{\rm T},\label{eq.phase_rock}
\end{eqnarray}
where $\Omega$ and $\phi_i^0$ are the constant frequency and the phase offset of the $i$th oscillator in the fully phase-locked solution.
Substituting Eq.~(\ref{eq.phase_rock}) into Eq.~(\ref{eq.phase_eq}), we obtain $\Omega$ explicitly as
\begin{eqnarray}
\Omega \equiv \omega_i+\sum_{j=1}^N \Gamma_{ij}^K (\phi_i^0-\phi_j^0).\label{eq.Omega_def}
\end{eqnarray}
By linearizing Eq.~(\ref{eq.phase_eq}) around the phase-locked solution Eq.~(\ref{eq.phase_rock}), 
we obtain: 
\begin{eqnarray}
\dot{\bm \psi}=L{\bm \psi},\label{eq.dynamics_around_phase_rock}
\end{eqnarray}
where ${\bm \psi} \equiv {\bm \phi}-{\bm \phi^*}$ is assumed to be small, and the elements $L_{ij}$ ($1 \le i, j \le N$) 
of the Jacobian $L$ of this system are given by 
\begin{eqnarray}
L_{ij}=\delta_{ij} \sum_{k \ne i}^{N} {\Gamma^K_{ik}}'(\phi_i^0-\phi_k^0)-(1-\delta_{ij}){\Gamma^K_{ij}}'(\phi_i^0-\phi_j^0).
\end{eqnarray}
Here, the prime denotes the derivative.
We solve the eigenvalue problem of this Jacobian:
\begin{eqnarray}
&&L{\bm u}^{(l)}=\lambda^{(l)} {\bm u}^{(l)},\label{eq.eigenvalue_right}\\
&&{\bm v}^{(l)} L=\lambda^{(l)} {\bm v}^{(l)},\label{eq.eigenvalue_left}
\end{eqnarray}
where $\lambda^{(l)}$, ${\bm u}^{(l)}=\{u_i^{(l)}\}$, and ${\bm v}^{(l)}=\{v_i^{(l)}\}$ 
are the $l$th eigenvalue and 
the $l$th right and left eigenvectors subject to the $l$th eigenvalue, respectively.
One of the eigenvalues, $\lambda^{(1)}$, is always zero owing to the rotational invariance 
${\bm \phi} \to {\bm \phi}+k {\bm u}^{(1)}$ in Eq.~(\ref{eq.phase_eq}) with $k$ being an arbitrary constant, 
where we define the corresponding right zero-eigenvector ${\bm u}^{(1)}$ as
\begin{eqnarray}
{\bm u}^{(1)}\equiv (1, 1, \cdots, 1)^{\rm T}.
\end{eqnarray}
Then, we obtain following relations for the left eigenvectors:
\begin{eqnarray}
\sum_{i=1}^{N} v_i^{(l)}=\delta_{l1},\label{eq.left_eigenvec_sum}
\end{eqnarray}
by using the orthonormality ${\bm v}^{(l)} {\bm u}^{(m)}=\delta_{lm}$.
Because of the assumption of the stability of the phase-locked solution, the real parts of the other eigenvalues are all negative
($0=\lambda^{(1)}> {\rm Re} \lambda^{(2)}\ge \cdots \ge {\rm Re}\lambda^{(N)}$).

We utilize the information of the left eigenvectors $v_i^{(l)}$ evaluated at $\hat{K}$, 
where $\hat{K}$ ($\ne K_{\rm c}$) is chosen from a range where the phase-locked solution exists. 
The value of $\hat{K}$ is generally different from the actual value of $K$ in Eq.~(\ref{eq.phase_eq}). 
Then, we consider the following nonlinear transformation of variables from 
${\bm \phi}$ to $\left(\Theta, R^{(1)}, \cdots, R^{(N)}\right)^{\rm T}$:
\begin{eqnarray}
&&R^{(1)}{\exp}(i \Theta) \equiv \sum_{i=1}^N \hat{v}_i^{(1)}
{\exp}({\rm i} (\phi_i-\hat{\phi}_i^0)),\label{eq.nonlin_transform1}\\ 
&&R^{(l)}\equiv \sum_{i=1}^N  \hat{v}_i^{(l)}{\exp}({\rm i} (\phi_i-\hat{\phi}_i^0-\Theta))\ \ (l\ne 1)\label{eq.nonlin_transform2},
\end{eqnarray}
where the quantities with the hat denote those evaluated at $K=\hat{K}$ hereafter.
The absolute value $R^{(1)}$ and the argument $\Theta$ are real variables. 
The argument $\Theta$ in Eq.~(\ref{eq.nonlin_transform2}) is defined in Eq.~(\ref{eq.nonlin_transform1}). 
The variables $R^{(l)}$ ($2\le l\le N$) are complex in general.
We call the variables $R^{(l)}$ nonlinear eigenmodes. This is motivated from the fact that,
as shown in step ({\rm B}), the linearization of $R^{(l)}$ reduces to the usual linear eigenmodes.
It may seem that the unknown degrees of freedom increase from $N$ to $2N$,
as the original real variables are transformed into complex ones.
However, because the norm of the quantity $\sum_{l=1}^{N} R^{(l)}\hat{u}_i^{(l)}=\exp({\rm i}(\phi_i-\hat{\phi}_i^0-\Theta))$ is preserved as
\begin{eqnarray}
\left| \sum_{l=1}^{N} R^{(l)}\hat{u}_i^{(l)} \right|=1,\label{eq.norm_conservation}
\end{eqnarray}
for all $i$, the number of the degrees of freedom are preserved under the transformation.
We also note that the transformation defined in Eqs.~(\ref{eq.nonlin_transform1}) and~(\ref{eq.nonlin_transform2}) 
has the important invariance under a $2\pi$-rotation of {\it each} oscillator $\phi_i \to \phi_i+2\pi$ for any $i$, as in the original equations in Eq.~(\ref{eq.phase_eq}). This feature becomes especially important when we apply our coarse-graining method to the system after a bifurcation of the phase-locked solution is exhibited.

\subsection{Dynamical equations for nonlinear eigenmodes}
By defining the new variables $\tilde{R}^{(l)}\equiv R^{(l)}\exp(i\Theta)$, 
we rewrite the phase equations in Eq.~(\ref{eq.phase_eq}) as
\begin{widetext}
\begin{eqnarray}
\dot{\tilde{R}}^{(l)}&&={\rm i}\sum_{j=1}^N \hat{v}_j^{(l)}\left(\sum_{k=1}^{M}\tilde{R}^{(k)}\hat{u}_j^{(k)}\right) 
\Biggl[\omega_j+\sum_{m=1}^{N}\Gamma^K_{jm} \Biggl \{{\rm arg}\left(\sum_{k=1}^{M}\tilde{R}^{(k)}\hat{u}_m^{(k)}\right)-{\rm arg} \left(\sum_{k=1}^{M}\tilde{R}^{(k)}\hat{u}_j^{(k)}\right)+\hat{\phi}_j^0-\hat{\phi}_m^0\Biggr\}\Biggr]\nonumber \\
&&+{\rm i} \epsilon \Biggl[ \sum_{j=1}^N \hat{v}_j^{(l)} \left(\sum_{k=1}^{M} \tilde{R}^{(k)}\hat{u}_j^{(k)} \right) Z\Biggl \{{\rm arg}  \left(\sum_{k=1}^{M} \tilde{R}^{(k)}\hat{u}_j^{(k)}\right)+\hat{\phi}_j^0\Biggr\}\xi_j(t) \Biggr] \ \ \ \ \ (l=1, \cdots, M),\label{eq.coarse_graining} 
\end{eqnarray}
\end{widetext}
where $M\le N$ is a constant. When $M=N$, Eq.~(\ref{eq.coarse_graining}) is equivalent to the original phase equations Eq.~(\ref{eq.phase_eq}).
We also consider the case $M < N$ when we perform the coarse-graining in step (C).
For $M=N$, the norm conservation in Eq.~(\ref{eq.norm_conservation}) also holds for $\tilde{R}^{(l)}$ as
\begin{eqnarray}
\left| \sum_{l=1}^{N} \tilde{R}^{(l)}\hat{u}_i^{(l)} \right|=1.\label{eq.norm_conservation2}
\end{eqnarray}
Then, Eq.~(\ref{eq.coarse_graining}) together with the norm conservation in Eq.~(\ref{eq.norm_conservation2}) constitute the dynamical equations for $\tilde{R}^{(l)}$. 

We note that by defining  
$\delta R^{(l)} \equiv \sum_{i=1}^N \hat{v}_i^{(l)}(\phi_i-(\Theta+\hat{\phi}_i^0))$ and neglecting the higher-order terms of $\delta R^{(l)}$,   
Eq.~(\ref{eq.coarse_graining}) with $M=N$ 
is rewritten as (see Appendix)
\begin{eqnarray}
&&\dot{\Theta}=\hat{\Omega}
,\label{eq.macro_phase1}\\
&&\delta \dot{R}^{(l)}=\hat{\lambda}^{(l)} \delta R^{(l)}+c \Delta K+\epsilon \sum_{j=1}^N \hat{v}_j^{(l)} Z(\Theta+\hat{\phi}_j^0)\xi_j(t)
,\label{eq.eigenmode}
\end{eqnarray}
where $\hat{\Omega} \equiv \omega_i+\sum_{j=1}^N \Gamma_{ij}^{\hat{K}} (\hat{\phi}_i^0-\hat{\phi}_j^0)$, $\hat{\lambda}^{(l)}\equiv \hat{{\bm v}}^{(l)}\hat{L} \hat{{\bm u}}^{(l)}$, $c={\rm const.}$, and $\Delta K\equiv K-\hat{K}$.
From Eq.~(\ref{eq.macro_phase1}), we obtain $\Theta=\hat{\Omega} t+\Theta_0$ with 
$\Theta_0$ being a constant. Without loss of generality, we choose $\Theta_0=0$ (i.e., $\Theta(t=0)=0$). 
Then, for $\Delta K=0$,
Eqs.~(\ref{eq.nonlin_transform1}) and (\ref{eq.nonlin_transform2}) reduce to a linear transformation from ${\bm \psi}$ to the usual linear eigenmodes ${\bm x}$ as $\delta R^{(l)}\simeq \hat{{\bm v}}^{(l)}{\bm \psi} \equiv x^{(l)}$ that diagonalize Eq.~(\ref{eq.dynamics_around_phase_rock}),
where we used the approximation $\Theta=\hat{\Omega}t$.
This is the reason why we call the variables $\tilde{{\bm R}}$ or ${\bm R}$ the nonlinear eigenmodes as a generalization of ${\bm x}$.

\subsection{Coarse-graining by neglecting unimportant eigenmodes}
For a phase-locked state with $\Delta K=0$ and $\epsilon=0$,  
we have $\tilde{R}^{(1)}=1$ and $\tilde{R}^{(l)}=0$ for $2 \le l\le N$, which can be confirmed by substituting the phase-locked solution 
${\bm \phi}^*$ in Eq.~(\ref{eq.phase_rock}) into the definitions in Eqs.~(\ref{eq.nonlin_transform1}) and (\ref{eq.nonlin_transform2}).
It is understandable that only the neutral eigenmode corresponding to the rotational invariance in a phase-locked state
becomes the unique effective degree of freedom. 

When the actual value of $K$ belongs to a range where the phase-locked solution does not exist (i.e., after the bifurcation) or even for a phase-locked state with $\Delta K\ne 0$, other effective eigenmodes also become effective in addition to the neutral eigenmode. 
Suppose that some $R^{(l)}$ ($l=N_{\rm r}+1, \cdots, N$) ($N_{\rm r} \ge 1$) 
are not relevant to the dynamics under consideration. 
In this case, we substitute $M=N_{\rm r}$ into Eq.~(\ref{eq.coarse_graining}) and assume
\begin{eqnarray}
\tilde{R}^{(l)}=0 \ \ (l=N_{\rm r}+1, \cdots, N).\label{eq.pro_coarse_graining}
\end{eqnarray}
Then, Eq.~(\ref{eq.coarse_graining}) is closed only by the effective eigenmodes $\tilde{R}^{(l)}$ ($l=1, \cdots, N_{\rm r}$).
This is the coarse-graining procedure of our method.
We note that $\tilde{{\bm R}}$ after the coarse-graining procedure in Eq.~(\ref{eq.pro_coarse_graining}) does not satisfy  
the norm conservation in Eq.~(\ref{eq.norm_conservation2}) anymore in a precise manner, whereas its violation may be small.  
 
We can approximately reproduce the original dynamics of ${\bm \phi}$ by inversely solving Eqs.~(\ref{eq.nonlin_transform1}) and (\ref{eq.nonlin_transform2}) as follows:
\begin{eqnarray}
\phi_i^{M}\equiv {\rm arg} \left(\sum_{l=1}^{M} \tilde{R}^{(l)} \hat{u}_i^{(l)}\right)+\hat{\phi}_i^0,\label{eq.realize_original_dynamics}
\end{eqnarray}
where ${\bm \phi}^M$ is the original ${\bm \phi}$ reproduced by Eq.~(\ref{eq.coarse_graining}).  
With ${\bm \phi}^M$, we can calculate various physical quantities.

\section{Example}
\begin{figure}[h]
\includegraphics[scale=0.85]{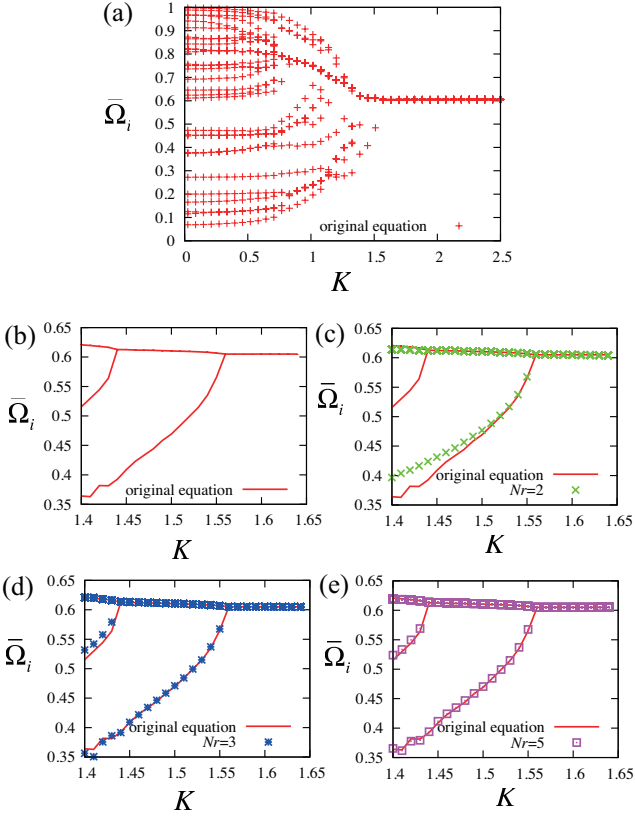}
\caption{(Color online) (a) $\bar{\Omega}_i$--$K$ diagram obtained numerically with the original phase equations in Eq.~(\ref{eq.original_random_net}). (b) Enlarged view of (a)
around the bifurcation of the phase-locked solution. 
(c)--(e) comparison of (b) with the diagram obtained numerically
with the coarse-grained equations in Eq.~(\ref{eq.coarse_graining}) with $M=N_{\rm r}=2, 3, 5$. 
}\label{fig.mean_freq}
\end{figure}
\begin{figure}[h]
\includegraphics[scale=0.75]{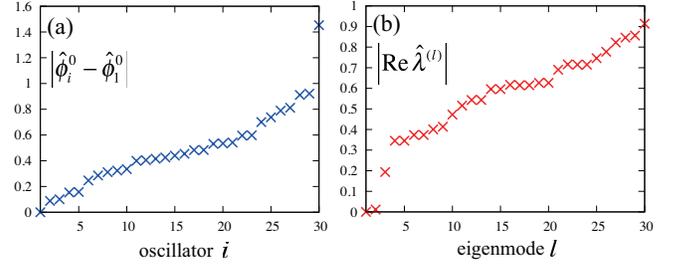}
\caption{(Color online) (a) The sorted absolute values of the relative phase offsets $\bigl|\hat{\phi}_{i}^0-\hat{\phi}_{1}^0\bigr|$ of
the fully phase-locked solution at $\hat{K}=1.555$, just before the bifurcation of the phase-locked solution.
The thirtieth oscillator $\phi_{30}$ is the oscillator that separates from the synchronized cluster of the other oscillators at the bifurcation point of the phase-locked solution, which is indicated in the gap 
between $\bigl|\hat{\phi}_{30}^0-\hat{\phi}_{1}^0\bigr|$ and $\bigl|\hat{\phi}_{29}^0-\hat{\phi}_{1}^0\bigr|$.
(b) The sorted norms of the real parts of the eigenvalues $\hat{\lambda}^{(l)}$s at $\hat{K}=1.555$, 
just before the bifurcation of the phase-locked solution. 
}
\label{fig.eigenvalue}
\end{figure}
To demonstrate our coarse-graining method, 
we apply it to the following phase equations on a random network as an example:
\begin{eqnarray}
\dot{\phi_i}=\omega_i+\frac{K}{N} \sum_{j=1}^{N}A_{ij}\sin(\phi_j-\phi_i),\label{eq.original_random_net}
\end{eqnarray}
where $K$ is the coupling strength, and $A_{ij}$ is the adjacent matrix given as 
\begin{eqnarray}
A_{ij}=
\begin{cases}
1, &  ({\rm probability} \ 0.5), \\
0, &  ({\rm otherwise}).
\end{cases} 
\end{eqnarray}
The adjacent matrix $A_{ij}$ is generally asymmetric.
The natural frequency $\omega_i$ for each oscillator is randomly sampled from the uniform distribution in the range $[0, 1]$. We use $N=30$.

As physical quantities to reproduce with our method, 
we consider the following time average of the frequency of each oscillator $\bar{\Omega}_i$ and the Kuramoto order parameter $\bar{r}$ defined as
\begin{eqnarray}
&& \bar{\Omega}_i \equiv \frac{1}{T} \int_0^T \dot{\phi}_i^M \ dt,\label{eq.omega_average}\\ 
&& r \equiv \frac{1}{N} \left|\sum_{i=1}^N \exp \left(-{\rm i}\phi_i^M\right)\right|, \ \bar{r}\equiv \frac{1}{T}\int_0^T r dt\label{eq.order_parameter},
\end{eqnarray}
respectively, where $T$ is a sufficiently long interval used for the average.
Using the fourth-order Runge-Kutta method with the time step $dt=0.001$, 
we first numerically solved the original phase equations in Eq.~(\ref{eq.original_random_net}) and 
measured the physical quantities in Eqs.~(\ref{eq.omega_average}) 
and (\ref{eq.order_parameter}) with $T=2400$.

In Fig.~\ref{fig.mean_freq} (a) [see also Fig.~\ref{fig.mean_freq} (b) for its enlarged view around the bifurcation point 
of the phase-locked solution], 
we plotted the $\bar{\Omega}_i$--$K$ diagram obtained with the original phase equations in Eq.~(\ref{eq.original_random_net}).
When $K$ is sufficiently large, we can see that the system is in a phase-locked state, and each oscillator oscillates with the shared frequency $\bar{\Omega}_i=\Omega$ given by Eq.~(\ref{eq.Omega_def}).
At the bifurcation point $K_{\rm c}$ ($1.554 <K_{\rm c} < 1.555$), the system exhibits a SN bifurcation, and the phase-locked solution disappears. Because of this SN bifurcation, one oscillator separates from the synchronized cluster of the other oscillators. 
[see also Fig.~\ref{fig.eigenvalue} (a) and its caption].
At the second bifurcation around $K=1.45$, another oscillator separates from the synchronized cluster.
As $K$ decreases further, the system exhibits a complicated cascade of bifurcations, and each oscillator oscillates independently at its natural frequency $\omega_i$ in the limit of $K \to +0$. 

Considering the result in Fig.~\ref{fig.mean_freq} (a), we choose $\hat{K}=1.555$, just before the bifurcation of the phase-locked solution, 
at which we solve the eigenvalue problem in Eqs.~(\ref{eq.eigenvalue_right}) and (\ref{eq.eigenvalue_left}). 
We first measured the phase offsets $\hat{\phi}_i^0$ around the fully phase-locked solution by the numerical simulation of Eq.~(\ref{eq.original_random_net}) [see Fig.~\ref{fig.eigenvalue} (a)].
The elements of the Jacobian of this system are given by $\hat{L}_{ij}=-\delta_{ij} \sum_{k \ne i}^{N} (\hat{K}/N)A_{ik}\cos(\hat{\phi}_i^0-\hat{\phi}_k^0)+(1-\delta_{ij})(\hat{K}/N)A_{ij} \cos(\hat{\phi}_i^0-\hat{\phi}_j^0)$. 
Then, we numerically solved the eigenvalue problem in Eqs.~(\ref{eq.eigenvalue_right}) and (\ref{eq.eigenvalue_left})
and performed the transformation in Eqs.~(\ref{eq.nonlin_transform1}) and (\ref{eq.nonlin_transform2}) with the obtained left eigenvectors. 

In Fig.~\ref{fig.eigenvalue} (b), we show the sorted absolute values of the real parts of the eigenvalues $\hat{\lambda}^{(l)}$ at $\hat{K}=1.555$.  
The eigenmode associated with $\hat{\lambda}^{(1)}=0$ corresponds to the neutral eigenmode resulting from the rotational invariance. 
The eigenmode associated with $\hat{\lambda}^{(2)}$ ($\bigl|{\rm Re}\hat{\lambda}^{(2)}\bigr| \ll 1$) corresponds to the bifurcation eigenmode at the bifurcation point $K=K_{\rm c}$.
This bifurcation eigenmode roughly expresses the relative motion between the synchronized cluster and the oscillator that separates from the cluster at the bifurcation of the phase-locked solution [see also Fig.~\ref{fig.eigenvalue} (a) and its caption].
The eigenmode associated with $\hat{\lambda}^{(3)}$ roughly expresses the relative motion between the synchronized cluster and 
the oscillator that separates from the cluster at the second bifurcation at around $K=1.45$.

\begin{figure}[t]
\includegraphics[scale=0.65]{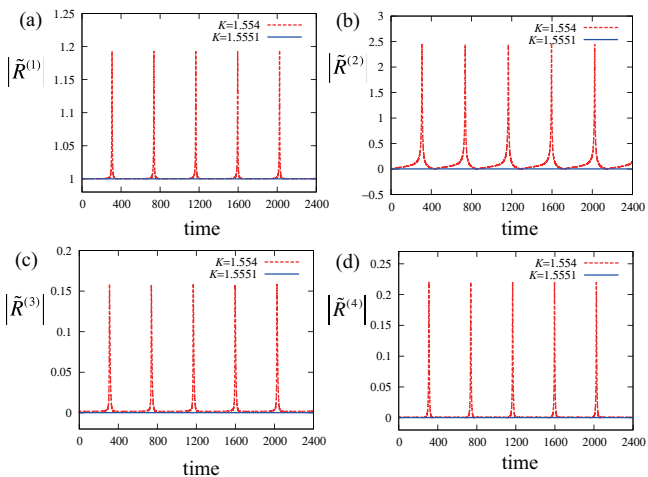}
\caption{(Color online) (a)--(d) Time evolution of the norms of some eigenmodes, $\tilde{R}^{(l)}$ $(l=1, 2, 3, 4)$,  
obtained by using Eqs.~(\ref{eq.coarse_graining}) with $M=N$ (no coarse-graining)
just before ($K=1.555$) and just after ($K=1.554$) the bifurcation of the phase-locked solution.
}\label{fig.R_zeroth}
\end{figure}
\begin{figure}[t]
\includegraphics[scale=1.0]{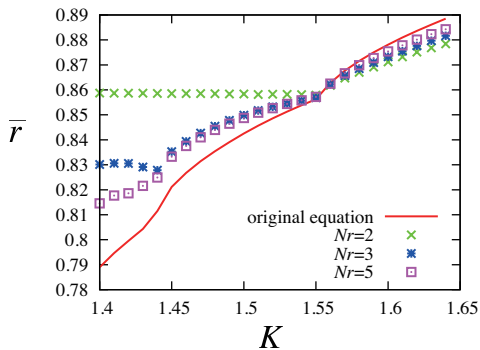}
\caption{(Color online)  
$\bar{r}$--$K$ diagram around the bifurcation of the phase-locked solution  
obtained numerically with the original phase equations in Eq.~(\ref{eq.original_random_net}) and with the coarse-grained equations in Eq.~(\ref{eq.coarse_graining}) with $M=N_{\rm r}=2, 3, 5$.
}\label{fig.kuramoto_para}
\end{figure}
In Figs.~\ref{fig.R_zeroth} (a)--(d), we show the typical time evolution of the norms of some eigenmodes,
$\bigl|\tilde{R}^{(l)}\bigr|$ ($l=1, 2, 3, 4$), 
obtained by using Eq.~(\ref{eq.coarse_graining}) with $M=N$ (no coarse-graining) just before ($K=1.555$) 
and just after ($K=1.554$) the bifurcation of the phase-locked solution.
Here, we also used the fourth-order Runge-Kutta method to solve Eq.~(\ref{eq.coarse_graining}) numerically 
by preparing $\tilde{R}^{(l)}=\delta_{1l}$ as an initial condition.
At $K=1.555$, i.e., $\Delta K=0$, just before the bifurcation, $\tilde{R}^{(l)}=\delta_{1l}$ continues to hold for all time.
At $K=1.554$, just after the bifurcation, we can see that each eigenmode remains close to the phase-locked solution almost all the time and sometimes exhibits oscillations with a large amplitude. 
This feature originates from the fact that the bifurcation of the phase-locked solution is the SN bifurcation; 
it accompanies an oscillation with a large amplitude and infinite period at the bifurcation point.

In Figs.~\ref{fig.mean_freq} (c)--(e), 
we compare the $\bar{\Omega}_i$--$K$ 
diagrams calculated by the original phase equations in Eq.~(\ref{eq.original_random_net}) and by the coarse-grained equations in Eq.~(\ref{eq.coarse_graining}) with $M=N_{\rm r}$.
We used $N_{\rm r}=2, 3, 5$ as the number of the effective eigenmodes.
With $N_{\rm r}=2$, we can reproduce the original diagram around the bifurcation of the phase-locked solution qualitatively as well as quantitatively.
With $N_{\rm r}=3$, we can also reproduce the original diagram even up to the second bifurcation occurring at around $K=1.45$.
This seems to be surprising because the second bifurcation is generally not related to the solution before the first bifurcation.
This seemingly nontrivial behavior may be explained by the following argument. 
First, the solutions before and after the first bifurcation are very similar because only one out of thirty oscillators is desynchronized at the first bifurcation.
Therefore, it is possible that the information regarding the second bifurcation is already obtained by the eigenvalue problem of the solution before the first bifurcation.

Finally, in Fig.~\ref{fig.kuramoto_para}, we plotted the $\bar{r}$--$K$ diagram.  
We can see that using $N_{\rm r}=2$ fails to reproduce the original diagram qualitatively, even in the vicinity of the bifurcation point of the phase-locked solution.
This is because using only two effective eigenmodes cannot express the complicated nonlinear dynamics inside the synchronized cluster consisting of twenty-nine oscillators with large-amplitude oscillations, 
which is inevitable in the SN bifurcation. 
In this sense, the minimum number of effective eigenmodes can be considered to be $3$.
By increasing the number of the effective eigenmodes, we would gradually reproduce the curvature of the original diagram
more precisely. 

\section{Discussion}
\subsection{Collective phase}
It is also interesting to consider the time evolution of a {\it collective phase}~\cite{RK,KKNAK,KNAKK,KNAKK2,KNAKK3,MKK}.
The collective phase is a single macroscopic phase constructed by the phases of oscillators, which may be observed by real experiments.
In~\cite{KKNAK}, the collective phase for a fully phase-locked state ${\Phi}$ 
is defined as ${\Phi} \equiv \sum_{i=1}^N \hat{v}_i^{(1)} (\phi_i-\hat{\phi}_i^0)$ (where we set $\Phi(t=0)=0$), and 
its time-evolution equation is derived as
\begin{eqnarray}
\dot{{\Phi}}=\hat{\Omega}+\epsilon \sum_{j=1}^N \hat{v}_j^{(1)}Z\bigl({\Phi}+\hat{\phi}_j^0\bigr)\xi_j(t)+O\left(\epsilon^2 \right).
\end{eqnarray}
As is clear from the definition of the collective phase, 
the left zero-eigenvector expresses how the microscopic phase of each oscillator contributes to the macroscopic phase.
The collective phase is related to $\Theta$ and $\delta R^{(1)}$ in our method as 
\begin{eqnarray}
{\Phi}=\Theta+\delta R^{(1)},
\end{eqnarray}
from their definitions. 
Thus, we can reproduce the time evolution of the collective phase $\dot{{\Phi}}$ as 
$\dot{{\Phi}}=\dot{\Theta}+\delta \dot{R}^{(1)}$ from Eqs.~(\ref{eq.macro_phase1}) and (\ref{eq.eigenmode}) with $\Delta K=0$ based on our framework. 
Therefore, our theory generalizes~\cite{KKNAK}.

As another intuitive definition of the collective phase, one may choose the center of the oscillators as ${\Phi}\equiv (1/N)\sum_{i=1}^N (\phi_i-\hat{\phi}_i^0)$.
A similar idea was also adopted in~\cite{RK},
and the time evolution of the collective phase by using the ``equation-free method'' was discussed.
In our method, this definition of the collective phase corresponds to the case where the Jacobian $\hat{L}$ is symmetric. 
In this case, because the left and right eigenvectors are proportional to each other,
$\hat{{\bm v}}^{(1)}=(1/N, 1/N, \cdots, 1/N)$ holds. 
Thus, the definition of the collective phase using the center of the oscillators is recovered. 
We note that our method can also be applied to systems with an asymmetric Jacobian without such restrictions.

\subsection{Comparison with other approximation methods} 
Here, we compare our coarse-graining method with other ones and clarify the characteristics and advantages of our method.
There are other various ways to eliminate ``fast variables'' 
and derive dynamical equations for ``slow variables'' closed only by them, 
which govern the essential behavior of the dynamical systems one considers. 
To be specific, let us consider the following time-evolution equations for the linear eigenmodes ${\bm x}$ 
around the bifurcation point:
\begin{eqnarray}
\dot{x}^{(l)}=\hat{\lambda}^{(l)}x^{(l)}+\sum_{m, k=2}^N a_{mk}^{(l)}x^{(m)}x^{(k)}+\cdots, \label{eq.dynamical_systems}
\end{eqnarray}
which can be derived by transforming the variables from ${\bm \phi}$ to $x^{(l)}=\hat{{\bm v}}^{(l)}{\bm \psi}$ in Eq.~(\ref{eq.phase_eq}) with $\epsilon=0$.
$a_{mk}^{(l)}$ denotes the expansion coefficients.
By this transformation, the neutral eigenmode $x^{(1)}$ does not appear in Eq.~(\ref{eq.dynamical_systems}).
The eigenmode associated with $\hat{\lambda}^{(2)}$ ($\bigl|{\rm Re} \hat{\lambda}^{(2)}\bigr| \ll 1$) 
corresponds to the bifurcation eigenmode that causes a SN bifurcation.
The adiabatic approximation and center manifold theory may be applied to Eq.~(\ref{eq.dynamical_systems}) 
for the derivation of the dynamical equations for slow variables. 

The adiabatic approximation formally eliminates the fast variables $x^{(l)}$ ($l=N_{\rm r}+1, \cdots, N$) by solving 
$\dot{x}^{(l)}=0$ and expressing these fast variables 
in terms of the slow variables $x^{(m)}$ ($m=2, \cdots, N_{\rm r}$) as 
$x^{(l)}=x^{(l)} \left(\bigl\{x^{(2)}, \cdots, x^{(N_{\rm r})}\bigr\}\right)$ ($l=N_{\rm r}+1, \cdots, N$). 
Then, we obtain the equations for the slow variables closed only by them by substituting these expressions for the fast variables into the equations for the slow variables. 
One of the difficulties of this method is that it is generally difficult to solve the complicated equations $\dot{x}^{(l)}=0$ ($l=N_{\rm r}+1, \cdots, N$) 
with many degrees of freedom.
Compared to the adiabatic approximation, 
our method is much easier to apply because 
we approximate the unimportant variables to be zero, as done in Eq.~(\ref{eq.pro_coarse_graining}).

The center manifold theory reduces higher-dimensional dynamical systems near a bifurcation to a one-dimensional system on the center manifold~\cite{SW}. 
The center manifold in this system is expressed as $x^{(l)}=b_2^{(l)} {x^{(2)}}^2+b_3^{(l)} {x^{(2)}}^3+\cdots$ 
($l=3, \cdots, N$), where $b_2^{(l)}, b_3^{(l)}, \cdots$ are the expansion coefficients that can be determined from Eq.~(\ref{eq.dynamical_systems}).
This is known as the center manifold theorem, and the reduced equation, which describes the time evolution of $x^{(2)}$ on the center manifold with ($N-2$) dimensions, is called the normal form.
The normal form describes the time evolution of $x^{(2)}$ exactly but only {\it locally} 
around the bifurcating solution.
In contrast, our method describes the time evolution of the eigenmodes {\it globally} but possibly only roughly, 
including the repeated oscillations with a large amplitude and long period  
due to the SN bifurcation, as shown in Figs.~\ref{fig.R_zeroth} (a)--(d).
This global feature results from the property in the transformation in Eqs.~(\ref{eq.nonlin_transform1}) and (\ref{eq.nonlin_transform2}); the transformation is invariant under the $2\pi$-rotation of each $\phi_i$, 
which is a property of the original phase equations in Eq.~(\ref{eq.phase_eq}). 
The normal form lacks such invariance in general.

\section{Summary}
In this paper, we proposed a coarse-graining method for general oscillator networks 
described by phase equations.
Our coarse-graining method enables us to obtain 
the closed coarse-grained equations for a few effective eigenmodes, 
based on the eigenvalue problem of the linearized system around the phase-locked solution and a nonlinear transformation.
We demonstrated our method by applying it to the phase equations of a random network,
which exhibits a SN bifurcation of the phase-locked solution at the bifurcation point. 
We emphasize the following advantages of our method. First, it can be applied to systems with an asymmetric Jacobian. 
Second, it can capture the global features of the dynamics, such as the one after the SN bifurcation 
that accompanies the oscillations with a large amplitude and long period, 
compared to the center manifold theory that describes only the local behavior around the bifurcating solution. 
Third, it is much easier to apply our method to the phase equations with many degrees of freedom rather than the adiabatic approximation. 
We expect that our method can provide a more brief description of the dynamics of the phase oscillators 
where only a few effective eigenmodes become essential. 

\begin{acknowledgements} 
The authors thank Fumito Mori for valuable discussions.
This work was supported by a Grant-in-Aid for Scientific Research on Innovative Areas ``The study on the neural dynamics for understanding communication in terms of complex hetero systems (No.4103)'' (13328715) of The Ministry of Education, Culture, Sports, Science, and Technology, Japan.
\end{acknowledgements} 

\appendix*
\section{Derivation of Eqs.~(\ref{eq.macro_phase1}) and (\ref{eq.eigenmode})}
By assuming that the quantity $\phi_j-(\Theta+\hat{\phi}_j^0)$ 
is sufficiently small and by using the definition Eqs.~(\ref{eq.nonlin_transform1}) and (\ref{eq.nonlin_transform2}), 
we can approximate $\tilde{{\bm R}}$ to
\begin{eqnarray}
\tilde{R}^{(l)}&&=\sum_{j=1}^N \hat{v}_j^{(l)} \exp \left({\rm i}(\phi_j-(\Theta+\phi_j^0)) \right) \exp \left({\rm i}\Theta \right)\nonumber\\
&&\simeq \sum_{j=1}^N \hat{v}_j^{(l)}(1+{\rm i}(\phi_j-(\Theta+\phi_j^0)))\exp \left({\rm i}\Theta \right)\nonumber\\
&&=(\delta_{1l}+{\rm i}\delta R^{(l)})\exp \left({\rm i}\Theta \right),\label{eq.R_approx}
\end{eqnarray}
where we used Eq.~(\ref{eq.left_eigenvec_sum}).
We also approximate several quantities in Eq.~(\ref{eq.coarse_graining}) with $M=N$ by using Eq.~(\ref{eq.R_approx}):
\begin{eqnarray}
&&\sum_{k=1}^N \tilde{R}^{(k)}\hat{u}_j^{(k)}\simeq \sum_{k=1}^N \left(\delta_{1k}+{\rm i}{\delta R}^{(k)} \right)\exp \left({\rm i}\Theta \right)\hat{u}_j^{(k)}\nonumber \\
&&=\Bigl(\underbrace{\sum_{k=1}^N \delta_{1k}\hat{u}_j^{(k)}}_{\hat{u}_j^{(1)}=1}+{\rm i} \sum_{k=1}^N \delta R^{(k)} \hat{u}_j^{(k)}\Bigr)\exp \left({\rm i}\Theta \right)\nonumber \\
&&=\Bigl(1+{\rm i} \sum_{k=1}^N \delta R^{(k)}\hat{u}_j^{(k)}\Bigr)\exp \left({\rm i}\Theta \right),\label{eq.approx1}
\end{eqnarray}
\begin{eqnarray}
&&{\rm arg}  \left(\sum_{k=1}^{N} \tilde{R}^{(k)}\hat{u}_j^{(k)}\right)=-{\rm i}\ln \left(\sum_{k=1}^N \tilde{R}^{(k)}\hat{u}_j^{(k)} \right)\nonumber \\
&&\simeq -{\rm i}\ln \Biggl\{\left(1+{\rm i} \sum_{k=1}^N \delta R^{(k)}\hat{u}_j^{(k)}\right)\exp \left({\rm i}\Theta \right)\Biggr\}\nonumber \\
&&\simeq \sum_{k=1}^N \delta R^{(k)}\hat{u}_j^{(k)}+\Theta,\label{eq.approx2}
\end{eqnarray}
\begin{widetext}
\begin{eqnarray}
&&{\rm i} \epsilon \Biggl[ \sum_{j=1}^N \hat{v}_j^{(l)} \left(\sum_{k=1}^{N} \tilde{R}^{(k)}\hat{u}_j^{(k)} \right) Z\Biggr\{{\rm arg}  \left(\sum_{k=1}^{N} \tilde{R}^{(k)}\hat{u}_j^{(k)}\right)+\hat{\phi}_j^0 \Biggr\}\xi_j(t) \Biggr]\nonumber \\
&&\simeq {\rm i}\epsilon \Biggl[ \sum_{l=1}^N \hat{v}_j^{(l)} \left(1+{\rm i} \sum_{k=1}^N \delta R^{(k)}\hat{u}_j^{(k)}\right) \Biggl\{Z\left(\Theta+\hat{\phi}_j^0 \right)+O(\delta R)\Biggr\}\xi_j(t) \Biggr]\exp \left({\rm i}\Theta \right)\nonumber\\
&&={\rm i} \epsilon \sum_{j=1}^{N} \hat{v}_j^{(l)} Z\left(\Theta+\hat{\phi}_j^0\right)\xi_j (t) \exp \left({\rm i}\Theta \right)+O\left( \epsilon\delta R \right).\label{eq.approx3}
\end{eqnarray}
\end{widetext}
Then, by substituting Eqs.~(\ref{eq.approx1}), (\ref{eq.approx2}), and (\ref{eq.approx3}) into Eq.~(\ref{eq.coarse_graining}) with $M=N$, 
we obtain the following relation up to
$O\left(\delta R, \epsilon, \Delta K \right)$ as
\begin{widetext}
\begin{eqnarray}
&&{\rm i}\delta_{1l}\dot{\Theta}\exp\left({\rm i}\Theta \right)+{\rm i}\delta \dot{R}^{(l)} \exp \left({\rm i}\Theta \right)+{\rm i}^2 \delta R^{(l)} \exp \left({\rm i}\Theta \right)\dot{\Theta}=
{\rm i} \sum_{j=1}^N \hat{v}_j^{(l)} \left(1+{\rm i} \sum_{k=1}^N \delta R^{(k)}\hat{u}_j^{(k)}\right)\exp \left({\rm i}\Theta \right)\times \nonumber \\
&& \underbrace{\Biggl\{\omega_j+\sum_{m=1}^N \Gamma^K_{jm} \left(\sum_{k=1}^N \delta R^{(k)}\hat{u}_m^{(k)}-\sum_{k=1}^N \delta R^{(k)} \hat{u}_j^{(k)}+\hat{\phi}_j^0-\hat{\phi}_m^0\right) \Biggr\}}_{\omega_j+\sum_{m=1}^N \Gamma^K_{jm} \left(\hat{\phi}_j^0-\hat{\phi}_m^0\right)+\sum_{m=1}^N {\Gamma^K_{jm}}' \left(\hat{\phi}_j^0-\hat{\phi}_m^0\right)\left(\sum_{k=1}^N \delta R^{(k)}\hat{u}_m^{(k)}-\sum_{k=1}^N \delta R^{(k)}\hat{u}_j^{(k)}\right)
+O\left(\delta R^2\right)}
+{\rm i} \epsilon \sum_{j=1}^{N} \hat{v}_j^{(l)} Z\left(\Theta+\hat{\phi}_j^0\right)\xi_j (t) \exp \left({\rm i}\Theta \right)\nonumber \\
&&={\rm i}\sum_{j=1}^{N} \hat{v}_j^{(l)}
\Biggl\{\underbrace{\omega_j+\sum_{m=1}^N \Gamma^K_{jm} \left(\hat{\phi}_j^0-\hat{\phi}_m^0\right)}_{\hat{\Omega}+O(\Delta K)}
+\sum_{i=1}^N\Bigl(\underbrace{\delta_{ji} \sum_{m\ne i}^N {\Gamma^K_{jm}}'\left(\hat{\phi}_j^0-\hat{\phi}_m^0\right)-(1-\delta_{ji}){\Gamma^K_{ji}}'\left(\hat{\phi}_j^0-\hat{\phi}_i^0\right)}_{\hat{L}_{ji}+O(\Delta K)}\Bigr)\left(\sum_{k=1}^N\hat{u}_i^{(k)}\delta R^{(k)}\right)\nonumber \\
&&+O\left(\delta R^2\right)\Biggr\}\times \exp \left({\rm i}\Theta \right)+{\rm i}^2\delta R^{(l)} \Biggl(\underbrace{\omega_j+\sum_{m=1}^N \Gamma^K_{jm}\left(\hat{\phi}_j^0-\hat{\phi}_m^0\right)}_{\hat{\Omega}+O(\Delta K)}+O(\delta R) \Biggr)\exp \left({\rm i}\Theta \right)\nonumber \\
&&+{\rm i} \epsilon \sum_{j=1}^{N} \hat{v}_j^{(l)} Z\left(\Theta+\hat{\phi}_j^0\right)\xi_j (t)\exp \left({\rm i}\Theta \right)\nonumber \\
&&={\rm i}\delta_{1l}\hat{\Omega} \exp \left({\rm i}\Theta \right)+{\rm i}\Bigl(c\Delta K+\sum_{k=1}^N\underbrace{\sum_{j, i=1}^N  \hat{v}_j^{(l)}\hat{L}_{ji} \hat{u}_i^{(k)}}_{\delta_{lk}\hat{\lambda}^{(k)}}\delta R^{(k)}+O\left(\delta R \Delta K, \delta R^2\right)\Bigr)\exp \left({\rm i}\Theta \right) \nonumber \\
&&+{\rm i}^2 \left(\delta R^{(l)}\hat{\Omega}+O(\delta R \Delta K, \delta R^2)\right)\exp \left({\rm i}\Theta \right)
+{\rm i} \epsilon \sum_{j=1}^{N} \hat{v}_j^{(l)} Z\left(\Theta+\hat{\phi}_j^0\right)\xi_j (t) \exp \left({\rm i}\Theta \right).\label{eq.coarse_graining_approx}
\end{eqnarray}
\end{widetext}
By comparing both sides of Eq.~(\ref{eq.coarse_graining_approx}), 
we obtain the time-evolution equations for $\Theta$ and $\delta R^{(l)}$ ($l=1, \cdots, N$) as
\begin{eqnarray}
&&\dot{\Theta}=\hat{\Omega}
,\label{eq.macro_phase}\\
&&\delta \dot{R}^{(l)}=
\hat{\lambda}^{(l)}\delta R^{(l)}+c\Delta K
+\epsilon \sum_{j=1}^N \hat{v}_j^{(l)} Z(\Theta+\hat{\phi}_j^0)\xi_j(t)
,\label{eq.linear_eigenmode}
\end{eqnarray}
respectively.


\begin{thebibliography}{9999}  
\bibitem{RK}K. Rajendran and I. G. Kevrekidis, Phys. Rev. E \textbf{84}, 036708 (2011).
\bibitem{BA} R. Albert and A. L. Barabasi, Rev. Mod. Phys. \textbf{74}, 47 (2002).
\bibitem{PRK} A. Pikovsky, M. Rosenblum, and J. Kurths, {\it Synchronization: A Universal Concept in Nonlinear Sciences}, (Cambridge University Press, Cambridge, 2001).
\bibitem{GFMS} T. Gregor, K. Fujimoto, N. Masaki, and S. Sawai, Science \textbf{328}, 1021 (2010).
\bibitem{W} A. T. Winfree, {\it The Geometry of Biological Time} (Springer, New York, 1980).
\bibitem{K} Y. Kuramoto, {\it Chemical Oscillations, Waves, and Turbulence} (Springer, New York, 1984).
\bibitem{KKNAK} H. Kori, Y. Kawamura, H. Nakao, K. Arai, and Y. Kuramoto, Phys. Rev. E \textbf{80}, 036207 (2009).
\bibitem{KNAKK} Y. Kawamura, H. Nakao, K. Arai, H. Kori, and Y. Kuramoto, Phys. Rev. Lett. \textbf{101}, 024101 (2008).
\bibitem{KNAKK2} Y. Kawamura, H. Nakao, K. Arai, H. Kori, and Y. Kuramoto, Chaos \textbf{20}, 043110 (2010).
\bibitem{KNAKK3} Y. Kawamura, H. Nakao, K. Arai, H. Kori, and Y. Kuramoto, Chaos \textbf{20}, 043109 (2010).
\bibitem{MKK} N. Masuda, Y. Kawamura, and H. Kori, New J. Phys. \textbf{12}, 093007 (2010).  
\bibitem{SW} S. Wiggins, {\it Introduction to Applied Nonlinear Dynamical Systems and Chaos}, 2nd ed. (Springer, New York, 2003).
\end{thebibliography}
\end{document}